\documentclass[oribibl,11pt]{llncs}
\usepackage{llncsdoc}

\usepackage{amsmath}
\usepackage{amssymb}
\usepackage{clrscode3e}
\usepackage{wrapfig}
\usepackage{graphics}
\usepackage{graphicx}
\usepackage{fullpage}
\usepackage{mathtools}
\usepackage{multirow}

\newcommand{\setGilt}[2]{\left\{ #1 \mid  #2\right\}}
\newcommand{\Is}       {:=}
\newcommand{\Id}[1]{\ensuremath{\text{{\sf #1}}}}
\newcommand{\set}[1]{\left\{ #1\right\}}
\newcommand{\ie}{i.e.\ }
\newcommand{\etal}{et~al.\ }

\newcounter{listcount} \newcounter{totcount}
\newcommand{\printarray}[2][1em]{% \printarray[<width>]{<array list>}
  \unskip \setcounter{totcount}{0}% Reset totcount counter
  \renewcommand*{\do}[1]{\stepcounter{totcount}}% Count elements
  \docsvlist{#2}% Process list a first time to obtain # of elements
  \setcounter{listcount}{0}% Reset listcount counter
  \renewcommand*{\do}[1]{%
    \stepcounter{listcount}% Move to next element
    \framebox[#1][c]{\rule{0pt}{1.9ex}\smash{\ensuremath{##1}}}%
    \ifnum\value{listcount}<\value{totcount}\thickspace\fi
  }
  \docsvlist{#2}% Process list a second time to typeset each element
}

\begin{document}
\title{(Semi-)External Algorithms for\\Graph Partitioning and Clustering}
\author{Yaroslav Akhremtsev, Peter Sanders and Christian Schulz\\ 
        \textit{Karlsruhe Institute of Technology (KIT)},
        \textit{Karlsruhe, Germany} \\
       \textit{Email: \{\protect\url{yaroslav.akhremtsev}, \protect\url{sanders}, \protect\url{christian.schulz}\}\protect\url{@kit.edu}} 
      }

\date{}

\institute{}

\maketitle
\begin{abstract}
In this paper, we develop semi-external and external memory algorithms for graph partitioning and clustering problems. 
Graph partitioning and clustering are key tools for processing and analyzing large complex networks.
We address both problems in the (semi-)external model by adapting the size-constrained \emph{label propagation} technique.
Our (semi-)external size-constrained label propagation algorithm can be used to compute graph clusterings and is a prerequisite for the (semi-)external graph partitioning algorithm.
The algorithm is then used for both the coarsening and the refinement phase of a multilevel algorithm to compute graph partitions.
Our algorithm is able to partition and cluster huge complex networks with billions of edges on cheap commodity machines.
Experiments demonstrate that the semi-external graph partitioning algorithm is scalable and can compute high quality partitions in time that is comparable to the running time of an efficient internal memory implementation.
A parallelization of the algorithm in the semi-external model further reduces running time.
\end{abstract}

\vfill
\pagebreak
\setcounter{page}{1}
\section{Introduction}
Graph partitioning and clustering problems are often solved to analyse or process large graphs in various contexts such as social networks, web graphs, or in scientific numeric simulations. 
To be able to process huge unstructured networks on cheap commodity machines one can rely on graph partitioning and partition the graph under consideration into a number of blocks such that each block fits into the internal memory of the machine while edges running between blocks are minimized (see for example \cite{kyrola2012graphchi}). 
However, to do so the partitioning algorithm itself has to be able to partition networks that do not fit into the internal memory of a machine.
In this paper, we present semi-external and external algorithms for the graph partitioning problem that are able to compute high quality solutions.

It is well known that graph partitioning is NP-complete~\cite{GareyJS74some} and that there is no constant factor approximation algorithm for general graphs~\cite{BuiJ92}. 
Hence, mostly heuristics are used in practice.
Probably the most-commonly used heuristic is the \emph{multilevel graph partitioning} (MGP) scheme. 
Here, the graph is recursively \emph{contracted} to obtain a sequence of smaller graphs with similar properties as the input graph. As soon as the graph is small enough, an \emph{initial partitioning algorithm} partitions the coarsest graph. Afterwards the contraction is undone and on each level a local search algorithm is used to improve the quality of the partition.

The paper is organized as follows. 
After introducing basic concepts and related work in Sections~\ref{s:preliminaries}, we present the key concepts and a rough overview of our technique in Section~\ref{s:multilevelgp}. 
Section~\ref{s:graphclusteringalgorithms} describes different algorithms to compute clusterings with or without a size-constraint in both computational models and a parallelization in the semi-external model.
Moreover, it presents the \emph{first} external memory algorithm to tackle the graph partitioning problem.
Subsequently, Section~\ref{s:se_e_multilevelgraphpartitioning} explains how clusterings can be used to build a graph hierarchy to be used in  a multilevel algorithm in the (semi-)external model.  
Experiments to evaluate the performance of our algorithms are presented in Section~\ref{s:experiements}. Finally, we conclude in Section~\ref{s:conclusion}.
In general, omitted proofs can be found in the appendix.

\section{Preliminaries}
\label{s:preliminaries}
\subsection{Basic Concepts} We consider an undirected weighted graph $G = (V, E,c,w)$,  with node weights $c: V \mapsto \mathbb{R}_{>0}$ and edge weights $w: E \mapsto \mathbb{R}_{>0}$. 
To keep our analysis simple, we assume that $|E| \geq |V|$. If the graphs are unweighted, we assume unit edge and node weights. 
The set $N(v) \Is \setGilt{u}{\set{v,u}\in E}$ denotes the neighbors of a node $v$.

Given a number $k>1$, the \emph{graph partitioning} problem asks to find blocks of nodes $V_1, \ldots, V_k$ such that $V_1 \cup \ldots \cup V_k = V$ and $\forall i \neq j: V_i \cap V_j = \emptyset$. 
A \emph{balance} constraint demands that $|V_i| \le L_{max}:= (1 + \varepsilon) \lceil|V|/k\rceil \ \forall i \in \{1, \ldots, k\}$, for some imbalance parameter $\varepsilon \geq 0$. 
The objective is to minimize the total \emph{cut} $\sum_{i<j}w(E_{ij})$ where 
$E_{ij}\Is\setGilt{\set{u,v}\in E}{u\in V_i,v\in V_j}$. 
Throughout the paper $\mbox{cluster}_G[v]$ denotes the block/cluster of a node $v$. We omit the subscript $G$ if the context allows it.
A \emph{graph clustering} is also a partition of the nodes, however, $k$ is not given in advance and there is no size-constraint. 
A size-constrained clustering constrains the size of the clusters by a given upper bound $U$ such that $|V_i| \leq U$. 
We say that a block/cluster $V_i$ is \emph{underloaded} if $|V_i| < L_{\max}$ and \emph{overloaded} if $|V_i| > L_{\max}$. 
Given a graph clustering $V_1, \ldots, V_k$, the \emph{quotient graph} is defined as $\mathcal{Q} = (V_q, E_q,c_q, w_q)$, where $V_q = \{1, \ldots, k\}$ and $E_q = \{(i, j) \mid E_{ij} \neq \emptyset\}$,~$w_q(i, j) = \sum_{e \in E_{ij}} w(e)$,~~$c_q(i) = \sum_{v \in V_i} c(v), ~i,j \in V_q$.

\paragraph{Computational Models.} We look at two models: the external and the semi-external model~\cite{journals/algorithmica/AbelloBW02}.
In both models, one wants to minimize the number of I/O operations.
In the external model it is assumed that the graph does not fit into internal memory whereas the semi-external memory assumes that there is enough memory for the nodes of the graph to fit into internal memory, but not enough for the edges.
We will use the following notations:
$M$ is the size of internal memory, $B$ is the size of a disk block, $\mathcal{O}(\frac{N}{B}) = \mbox{Scan}(N)$ is the number of I/O operations needed for reading or writing an array of size $N$ and $\mathcal{O}(\frac{N}{B}  \log_{\frac{M}{B}}\frac{N}{B}) = \mbox{Sort}(N)$ is the number of I/O operations needed for sorting an array of size $N$.

\paragraph{Graph Data Structure.}
To store a graph in external memory, we use a data structure similar to an adjacency array. 
This data structure allows us to inspect all edges using $\mbox{Scan}(|E|)$ I/O operations.
An external array of the edges contains the adjacency lists of each node in increasing order of their IDs. 
Each element of the adjacency list of a node $u$ is a pair $(v, w)$, where $v$ is the target of the edge $(u, v)$ and $w = w(u, v)$ is the weight of the edge. 
We mark the end of each adjacency list by using a sentinel pair. This allows us to determine easily if we reached the end of the adjacency list of the node that we currently process.
The second external array stores node offsets, \ie for each node we store a pointer to the beginning of its adjacency list in the edge array. 
The third external array contains the weights of the nodes.

\subsection{Related Work}
There has been a \emph{huge} amount of research on graph partitioning so that we refer the reader to \cite{GPOverviewBook,SPPGPOverviewPaper} for most of the material. 
Here, we focus on issues closely related to our main contributions. 
All general-purpose methods that work well on large 
real-world graphs are based on the multilevel principle. 
The basic idea can be traced back to multigrid
solvers for systems of linear equations.
Recent practical methods are mostly based on graph theoretic aspects, in
particular edge contraction and local search.  
There are different ways to create graph hierarchies such as matching-based schemes (e.g. \cite{diekmann2000shape}) or variations thereof~\cite{Karypis06} and techniques similar to algebraic multigrid (e.g. \cite{meyerhenke2006accelerating}). 
Well-known MGP software packages include Chaco~\cite{chaco}, Jostle~\cite{walshaw2000mpm}, Metis~\cite{karypis1998fast}, Party~\cite{helpfulsetsinpractice} and Scotch~\cite{Scotch}.   

Graph clustering with the label propagation algorithm (LPA) has originally been described by Raghavan
\etal \cite{labelpropagationclustering}.
Meyerhenke \etal \cite{pcomplexnetworksviacluster} introduced a size-constrained LPA. 
Doing so made it possible to use this algorithm during coarsening and uncoarsening of a multilevel scheme to compute graph partitions of large complex networks. 
In this paper, we present a semi-external and an external variant of this algorithm.
There are other semi-external algorithms to tackle the graph partitioning problem which are based on streaming~\cite{Stanton:2012:SGP:2339530.2339722}. 
However, they do not use a multilevel approach and do not achieve high solution quality.
To the best of our knowledge, our algorithm is the first that tackles the graph partitioning problem in the external memory model.

\paragraph{KaHIP}
\label{s:kaHIP}
(Karlsruhe High Quality Partitioning) is a family of graph partitioning algorithms that tackle the balanced graph partitioning problem~\cite{kabapeE}.  
It includes several multilevel algorithms and meta-heuristics to compute high quality partitions.
In particular, the algorithms of Meyerhenke \etal \cite{pcomplexnetworksviacluster} to partition large complex networks are included.
We use KaHIP to partition the coarse graphs as soon as they fit into the internal memory of the machine.

\section{Basic Cluster Contraction}
\label{s:multilevelgp}
We now present the basic idea and explain the main ingredients which are used to obtain a (semi-)external graph partitioning algorithm. 
First of all, we define a (semi-)external algorithm to create graph hierarchies for the multilevel graph partition approach. 
The approach of Meyerhenke \etal \cite{pcomplexnetworksviacluster} to create graph hierarchies in internal memory iteratively contracts size-constraint graph clusterings that are obtained using a \emph{label propagation algorithm}.
\emph{Contracting a clustering} works as follows: 
each block of the clustering is contracted into a single node. 
The weight of the node is set to the sum of the weight of all nodes in the original block. 
There is an edge between two nodes $u$ and $v$ in the contracted graph if the
two corresponding blocks in the clustering are adjacent to each other in $G$,
\ie block $u$ and block $v$ are connected by at least one edge.
The weight of an edge $(A,B)$ is set to the sum of the weight of edges that run between block $A$ and block $B$ of the clustering. 
Due to the way contraction is defined, a partition of the coarse graph corresponds to a partition of the finer graph with the same cut and balance. Note that the contracted graph corresponds to the quotient graph.

Cluster contraction is an aggressive coarsening strategy. In contrast to most previous approaches, it can drastically shrink the size of irregular networks.
The intuition behind this technique is that a clustering of the graph (one hopes) contains many edges running inside the clusters and only a few edges running between clusters, which is favorable for the edge cut objective.
Regarding complexity, experiments in \cite{pcomplexnetworksviacluster} indicate that already one contraction step can shrink the graph size by orders of magnitude and that the average degree of the contracted graph is smaller than the average degree of the input network. 
Thus it is very likely that the graph will fit into internal memory after the first contraction step.
On the other hand, by using a different size-constraint ($|V_i| \leq L_\text{max}$), the LPA can also be used as a simple strategy to improve a solution on the current level. 

Our \emph{main idea} to obtain a (semi-)external multilevel graph partitioning algorithm is to engineer (semi-)external variants of the size-constrained LPA and to externalize the contraction as well as solution transfer component of the algorithm. 
By doing this, we have a (semi-)external algorithm to build graph hierarchies and to transfer solutions to finer levels. 
Once the graph is small enough to fit into internal memory, we use the KaHIP framework to compute a partition of the graph.
Additionally, we use a (semi-)external size-constrained LPA as a local search algorithm to improve the solution on the finer levels that do not fit into internal memory.
We proceed by explaining the (semi-)external size-constrained LPA as well as its parallelization in Section~\ref{s:graphclusteringalgorithms} and explain the external contraction and solution transfer algorithm in Section~\ref{s:se_e_multilevelgraphpartitioning}.

\section{(Semi-)External Graph Clustering}
\label{s:graphclusteringalgorithms}
We now explain how graph clusterings can be obtained in both, the semi-external and the external memory model.
We present multiple algorithms: a semi-external LPA that can deal with size-constraints, an external LPA that does not use size-constraints, as well as a coloring-based graph clustering algorithm inspired by label propagation that can also maintain size-constraints in the external model. 

\subsection{Label Propagation}
\emph{Label propagation} (LP) was first presented by Raghavan \etal \cite{labelpropagationclustering}. 
In the beginning each node belongs to its own cluster. 
Afterwards the algorithm works in rounds. In each round, the algorithm visits each node in increasing order of their IDs.  
When a node $v$ is visited, it is \emph{moved} to the block that has the strongest connection to $v$, \ie it is moved to the cluster $V_i$ that maximizes $\omega(\{(v, u) \mid u \in N(v) \cap V_i \})$. 
Ties are broken randomly. 
If the algorithm is used to compute a size-constrained clustering, the selection rule is modified such that only moves are eligible that do not result in overloaded blocks.

Suppose the algorithm is currently processing a node $v$. 
Now we scan the adjacency list of the node $v$ in the respective memory model and compute the new cluster of this node.
If a size-constraint is present, we also need a scheme to manage the sizes of each cluster/block.

\paragraph{Semi-External Label Propagation.} This is the simple case: since we have $\mathcal{O}(|V|)$ internal memory, we can afford to store the cluster IDs in internal memory.
Additionally, we maintain an array of size $|V|$ in internal memory that stores the cluster sizes. Hence, one iteration of the semi-external LP algorithm can be done using $\mbox{Scan}(|E|)$ I/O operations.

\begin{figure}[h]

\vspace*{-.5cm}
\begin{center}
\includegraphics[scale=0.75]{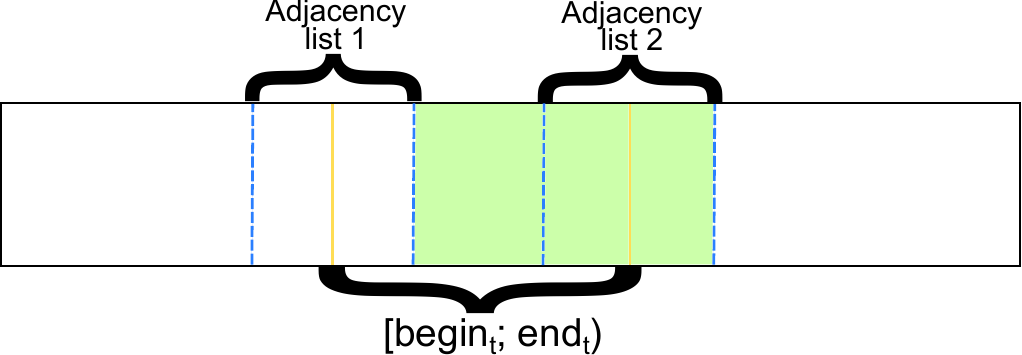}
\caption{Range processed by a thread}
\label{fig:block_range}
\end{center}
\vspace*{-1.cm}
\end{figure}

\paragraph{Parallel Semi-External Label Propagation.} Recall that the LP algorithm iterates through the external array of edges. In attempt to accelerate semi-external LPA and to get closer to the I/O bound, we parallelize the processing of a disk block of edges. 
Since, the LP algorithm processes nodes interdependently, we divide the disk block into equal ranges and process them in parallel. 
We now explain how we process nodes which have adjacency lists that belong to different disk blocks (see Figure~\ref{fig:block_range} for an example). 
Each thread $t$ begins to process its range ${[}begin_t; end_t)$ of the disk block.
Afterwards, the range is shifted such that each adjacency list in the block is processed by precisely one thread.
Consider the example depicted in Figure~\ref{fig:block_range}. 
Here, the thread finds the end of the adjacency list 1 in ${[}begin_t; end_t)$ and iterates through the elements until the end of adjacency list 2 is reached. 
The colored area in Figure~\ref{fig:block_range} represents the range that will be actually processed by thread $t$.

To maintain up-to-date cluster sizes, we do not move a node immediately, i.e. we store the moves which were generated by the threads during the processing of the disk block. Afterwards, all moves are processed sequentially and we make a move if it does not violate the size constraint. 

\paragraph{External Label Propagation.} To propagate the cluster IDs of adjacent nodes, we use time forward processing \cite{ZehIO}.
More precisely, we maintain two external priority queues~\cite{Sanders99fastpriority}: one for the current and one for the next round. Initially, the current queue contains triples $(v, c, w)$ for each edge $(u, v) \in E: v < u$ where $v$ is the key value, $w$ denotes the weight of the edge and $c$ is the current cluster ID of $u$. 
When the algorithm scans a node $u$, all triples $(u, c, w)$ are on top of the  priority queue since the nodes are processed in increasing order of their ID.
The tuples are then extracted using the operations $\proc{Pop}$ and $\proc{Top}$. 
This means, we know the current cluster ID of all adjacent nodes of $u$ and can calculate the new cluster ID. 
When the new cluster ID is computed, the algorithm pushes triples with the new cluster ID for all adjacent nodes into the next and current priority queue depending on the node ID of the neighbor $v$: if $u < v$ we push $(v, \mbox{cluster}[u], w(u, v))$ to the current priority queue and if $v < u$ we push it to the next priority queue.
At the end of a round we swap the priority queues.

Each operation of the priority queue $\proc{Pop}$, $\proc{Push}$ and $\proc{Top}$ is called $\mathcal{O}(|E|)$ times. Each of the operations can be performed using $\mathcal{O}(\frac{1}{B}\log_{\frac{M}{B}}{\frac{|E|}{B}})$ I/O operations amortized~\cite{Sanders99fastpriority, journals/algorithmica/Arge03}. Thus, the overall algorithm uses $\mathcal{O}(\frac{|E|}{B}\log_{\frac{M}{B}}{\frac{|E|}{B}}) = \mbox{Sort}(|E|)$ I/O operations.

\subsection{Coloring-based Graph Clustering}
We now present another approach to compute a graph clustering in the external model. 
The algorithm is able to maintain the sizes of all clusters in the external memory model.
The \emph{main idea} of the algorithm is to process \emph{independent sets}. 
Due to the definition of an independent set, a change of the cluster ID of a node will not have an effect on the other nodes within the set.
However, the changes of all adjacent nodes of $v$ need to be taken into account. 

Assume that we have a node coloring $C = \{C_1, C_2, \ldots, C_\ell\}$ of the graph, where $C_i$ is the set of vertices with the same color $i$. 
Note that each set $C_i$ forms an independent set.  For each set $C_i$, we maintain an external array (bucket) of tuples $\mathcal{T}_i$ 
and allocate a buffer of size $B$ internal memory for each bucket $\mathcal{T}_i$. Hence, we assume that the number of colors $|C|$ is smaller than $\frac{M}{B}$.  

The bucket clustering algorithm also works in rounds. 
Roughly speaking, in each round it processes the buckets in increasing order of their color and updates the cluster IDs of all nodes.
When we process a bucket we need the cluster IDs of all adjacent nodes.
Hence, we define the content of a bucket as follows:
we store tuples so that we know  the current cluster ID of each adjacent neighbor, its color, ID of the neighbor and the weight of the edge.  
More precisely,  initially for a node $u$, we store the following tuples for all adjacent nodes $v$ with $\mbox{color}[v] < \mbox{color}[u]$ in the corresponding buckets $\mathcal{T}_{\mbox{color}[v]}$: $(v, \mbox{cluster}[u], u, w(u, v), \mbox{color}[u])$. To do this efficiently,  
we augment the array of edges by adding the color of the target node $v$ to each edge $(u,v)$ before the initialization step.
Note that these tuples contain the complete information about the graph structure and that the information suffices to update the cluster of a node.

When the algorithm processes a bucket $\mathcal{T}_i$, it sorts the elements of the bucket lexicographically by the first and second component. 
Afterwards, it scans the tuples of the bucket and calculates a new cluster ID for each node in $C_i$ in the same manner as the LPA. 
When the bucket is processed, we push tuples with the new cluster IDs to the corresponding bucket, \ie for each tuple $(v, \mbox{cluster}[u], u, w(v, u), \mbox{color}[u])$ in bucket $\mathcal{T}_i$, we push the tuple $(u, \mbox{cluster}[v], v, w(u, v), \mbox{color}[v])$ into the bucket $\mathcal{T}_{\mbox{color}[u]}$. 

\begin{lemma}
\label{th:bucket_aux}
Processing a bucket $\mathcal{T}$ requires $\mbox{Sort}(|\mathcal{T}|)$ I/O-operations. 
\end{lemma}

\begin{theorem} 
\label{th:bucket_main}
The bucket algorithm requires $\mbox{Sort}(|E|)$ I/O-operations for one iteration of label propagation. 
\end{theorem}

\paragraph{Graph Coloring.}
Computing the graph coloring is a very important part of the bucket graph clustering algorithm. Note that the number of colors is equal to the number of buckets and we want to maintain as few buckets as possible. This is due to the fact that we need an amount of $B$ space for each bucket in internal memory. 
Moreover, the size of each bucket must be smaller than an upper bound, since each bucket has to fit into internal memory during our experiments. 
To compute a coloring, we use the time forward processing technique~\cite{ZehIO} with an additional size-constraint on the color classes that can be maintained in internal memory. This allows us to build a coloring using $\mbox{Sort}(|E|)$ I/O operations. Note that the coloring is computed only once so that the cost
for computing the coloring can be amortized over many iterations of label propagation.

\subsubsection{External Graph Clustering Algorithm with Size-Constraints.} 
In this section we describe how we  modify the coloring-based clustering algorithm, so that it can handle a size-constraint. 
The main advantage of the coloring-based clustering algorithm is as follows.
When we process a bucket, the cluster IDs of all adjacent nodes will not change.
This allows us to maintain a data structure with up-to-date sizes of the clusters of the nodes of the independent set and their neighbors.
In the following, we consider two different data structures depending on if each of the buckets fits into internal memory or not.
In both cases, we use an external array that stores the sizes of all clusters. We start by explaining the case where each bucket fits into internal memory.

\paragraph{Each bucket fits into internal memory.} In this case, we can use a hash table $\mathcal{H}$ to maintain the cluster sizes of the current bucket. 
The key of $\mathcal{H}$ is the cluster ID and the value is the current size of the cluster. 
When we process a bucket $\mathcal{T}_i$,  the hash table $\mathcal{H}$ can be built as follows. 
We collect all cluster IDs of the nodes of the current independent set as well as their neighbors, sort them and then iterate through the external array to get the current cluster sizes. 
After finishing to calculate a new cluster ID for each node in $C_i$, we write the updated cluster sizes to the external array. Hence, the cluster sizes are up-to-date after we processed the current bucket. 

\begin{theorem}
\label{th:bucket_sz}
The coloring-based clustering algorithm with size-constraints uses $t \cdot \mbox{Scan}(|V|) + \mbox{Sort}(|E|)$ I/O operations, where $t = \max(\frac{|E|}{M}, |C|)$ is the amount of buckets such that each bucket fits into internal memory.
\end{theorem}

\paragraph{There is at least one bucket that does not fit into internal memory.} 
This case is somewhat more complicated, since we cannot afford to store the hash table in internal memory.
Basically, when we process a bucket $\mathcal{T}_i$, we do not use a hash table but an external priority queue and additional data structures which contain enough information to manage the cluster sizes. 
More precisely, we define a structure $\mathcal{M}$ that tells us for each node which nodes need the updated cluster size information.
Nodes from $C_i$ are still processed in increasing order of their IDs. We now explain the structures in detail.

\begin{wrapfigure}{l}{.4\textwidth}
\centering
\vspace*{-.5cm}
\includegraphics[scale=0.35]{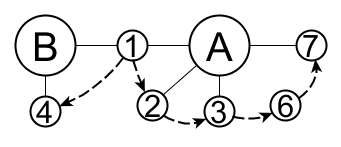}
\caption[]{In this example, we have $\mathit{C}(1) = \{A, B, \mbox{cluster}[1]\}$ and $\mathit{C}(4) = \{B, \mbox{cluster}[4]\}$. Dotted lines denote forwarding cluster size changes. A, B are cluster IDs, 1-7 denote node ID. Node 6 belongs to cluster A. 
The sets of adjacent nodes for the cluster $A$ and $B$ are  $\mathit{N}_A = \{1, 2, 3, 6, 7\}$ and $\mathit{N}_B = \{1, 4\}$.
Moreover, $\mathit{M}_1 = \{(2, A), (4, B)\}$, $\mathit{M}_2 = \{(3, A)\}$.
}
\label{fig:external_graph_partition_0}
\vspace*{-.5cm}
\end{wrapfigure}

For a node $v$ in the current independent set $C_i$, let $\mathit{C}(v) := \{\mbox{cluster}[u] \mid (v, u) \in E \} \cup \{\mbox{cluster}[v]\}$ denote the set of adjacent clusters. 
An example is shown in Figure~\ref{fig:external_graph_partition_0}. 
These are the clusters that can possibly change their size if $v$ changes its cluster. 
We now need to find all nodes from the independent set that are adjacent to the clusters or are contained in this  cluster because they need to receive 
the updated cluster size.

The first additional data structure contains only nodes of the independent set. It is needed to build the next data structure $\mathcal{M}$.
Let $\mathit{N}_c:= \{u \in C_i \mid \exists (u, v) \in E :   \mbox{cluster}[v] = c\} \cup \{u \in C_i \mid \mbox{cluster[$u$]} = c\}$ be the set of adjacent nodes for a cluster $c$ that are in the current independent set (including the nodes that are in the cluster). 
We sort $\mathit{N}_c$ in increasing order of node IDs  and remove repeated elements. 
For a cluster $c$, the set $N_c$ contains all nodes from the independent set that are adjacent to the cluster. 
Moreover, the order in $N_c$ is similar to the processing order of the independent set.
We denote the $j$-th node of $\mathit{N}_c$ as $\mathit{N}_c^j$. 
The second additional data structure uses the first one and is defined as the set $\mathit{M}_v:= \{(u, c) \mid  c \in C(v),\  \mathit{N}_c^j = v,\ \mathit{N}_c^{j + 1} = u\}$. 
It contains the nodes to which the node $v$ must forward information about the changes in the cluster sizes. 
Roughly speaking, for each cluster in the neighborhood of $v$ (including the cluster of $v$), $M_v$ contains the adjacent node of the cluster that will be processed next.
This way the information can be propagated easily.
An example is shown in  Figure~\ref{fig:external_graph_partition_0}.
\ \ \ We now explain the details of the algorithm when processing one bucket.
First, we compute the sets $\mathit{N}_c$. 
To do so, we build a list $N$ of pairs $(c, v)$ that are sorted lexicographically by their first and second component, where $c$ is the ID of the cluster adjacent to $v$. For building this list, we iterate through the bucket and add pairs $(c, v)$ for each tuple $(v, \ldots) \in \mathcal{T}_i$ to the list and also add the pair $(\mbox{cluster}[v], v)~ \forall v \in C_i$. Then we sort these pairs and we are done. 
Note that  $|\mathit{N}| = |C_i| + |\mathcal{T}_i| = \mathcal{O}(|\mathcal{T}_i|)$.
To compute the sets $\mathit{M}_v$, we build a list $\mathcal{M}$ of triples $(v, c, u)$, where $v$ is the node ID and $(u, c) \in \mathit{M}_v$. For each $\mathit{N}_c^j = v$ and $\mathit{N}_c^{j + 1} = u$ the triple $(v, c, u)$ is added to the list. Afterwards, the triples are sorted by the first component. 
Note that the size of the list is at most $\mathcal{O}(|\mathcal{T}_i|)$.

Recall, that the set $M_v$ contains the nodes that have to receive the changes in the cluster sizes.
To forward the information, we use an external priority queue. 
The priority queue contains triples $(v, c, sz)$, where $v$ is the node ID which also serves as key value, $c \in C(v)$ is the cluster and $sz$ the size of the cluster. 
We initialize the priority queue as follows: we iterate through the sets $\mathit{N}_c$ and put the tuples $(v_1, c, sz)$ in the priority queue, where $v_1$ is the first node in $\mathit{N}_c$. The sizes of the clusters are obtained from the external array containing the cluster sizes. 
Then the nodes are processed. 
After node $v$ is processed, we put $(u, c, sz)$ in the priority queue for each pair $(u, c) \in \mathit{M}_v$. 
After we processed a bucket, we update the cluster  sizes in the external array.

\begin{lemma}
\label{th:bucket_blubb}
When node $v$ is processed there is a triple $(v, c, sz)$ for each adjacent cluster on the top of priority queue with up-to-date cluster sizes.
\end{lemma}

\begin{lemma}
\label{th:bucket_lemma_sz_ext}
Processing a bucket and maintaining cluster sizes costs $\mbox{Sort}(|\mathcal{T}|)$ I/O-operations.
\end{lemma}

\begin{theorem}
\label{th:bucket_th_sz_ext}
One iteration of the coloring-based clustering algorithm costs $\mbox{Sort}(|E|) + |C| \cdot \mbox{Scan}(|V|)$ operations, where $|C|$ is the number of buckets.
\end{theorem}

\section{(Semi-)External Multilevel Graph Partitioning}
\label{s:se_e_multilevelgraphpartitioning}
We now explain the (semi-)external multilevel graph partitioning algorithm. 
\paragraph{Coarsening/Contraction.} We have two different algorithms to create graph hierarchies depending on the memory model that we use. 
In general to create a graph hierarchy, we compute a clustering with size-constraints of the current graph using some algorithm from Section~\ref{s:graphclusteringalgorithms}.   The next step is to renumber the cluster IDs. 
The external algorithm sorts the nodes by their cluster ID and scans the sorted array assigning new cluster IDs from $0, \ldots, n' - 1$, where $n'$ is the number of the distinct clusters. 
This step can be done using $\mbox{Sort}(|V|)$ I/O operations. 
In contrast, the semi-external algorithm uses an additional array of size $\mathcal{O}(|V|)$ to assign new cluster IDs. Hence, it needs $\mathcal{O}(1)$ I/O operations.

The external algorithm builds an array of triples $(\mbox{cluster}[u], \mbox{cluster}[v], w(u, v))$ for each edge $(u, v) \in E$ to build the contracted graph. 
This array is sorted lexicographically by the first two entries using $\mbox{Sort}(|E|)$ I/Os. 
Then we merge parallel edges and build the edges of the quotient graph by iterating through the sorted array using $\mbox{Scan}(|E|)$ I/Os. 
The total I/O volume of this step is $\mbox{Sort}(|E|)$. 
The semi-external algorithm stores pairs $(\mbox{cluster}[u], \mbox{cluster}[v])$ for each edge $(u, v)$ in a hash table and uses it to build the contracted graph. 
This can be done using $\mbox{Scan}(|E|)$ I/Os.
If the number of edges of the contracted graphs decreases geometrically and a constant number of label propagation iterations is assumed, the complete hierarchy can be built using $\mbox{Scan}(|E|)$ I/Os using the semi-external algorithm or $\mbox{Sort}(|E|)$ I/Os using the external algorithm.

\paragraph{Uncoarsening/Solution Transfer.}
In this step, we want to transfer a solution of a coarse level to the next finer level in the hierarchy and perform some local search.
Let $\mathcal{Q} = (V_{\mathcal{Q}}, E_{\mathcal{Q}})$ be a contracted graph of the next finer level $G=(V, E)$ and let $\mbox{cluster}_\mathcal{Q}$ be a partition of the contracted graph.  
Recall that the contracted graph has been built according to a clustering of the graph $G$. 
Also note that cluster $i$ of $G$ corresponds to node $i$ in the contracted graph. Hence, for a node $v \in V$ the transferred cluster ID of the coarse level is $\mbox{cluster}_G'[v]:= \mbox{cluster}_{\mathcal{Q}}[\mbox{cluster}_G[v]]$.  

To transfer the solution in external algorithm, we build an array of pairs $(\mbox{cluster}_G[v], v)$ and sort it by the first component using $\mbox{Sort}(|V|)$ I/O operations. Now we iterate through both of the arrays $\mbox{cluster}_{\mathcal{Q}}$ and  $\{(\mbox{cluster}_G[v], v)\}$ at the same time and generate the array $\{(\mbox{cluster}_{\mathcal{Q}}[\mbox{cluster}_G[v]], v)\}$ which contains the transferred solution. 
We sort the resulting array by the second component and apply the clustering to our graph.
Overall, we need $\mbox{Sort}(|V|)$ I/O operations. 
The semi-external algorithm iterates through all nodes of graph $G$ and updates the cluster IDs of each node. This can be done using $\mathcal{O}(1)$ I/Os.
If the number of nodes of the quotient graphs decreases geometrically then uncoarsening of the complete hierarchy can be done using $\mbox{Sort}(|V|)$ I/O operations.
After each solution transfer step, we apply a size-constrained LPA (using $L_\text{max}$) to improve the solution in (semi-)external memory on the current level.

\section{Experiments}
\label{s:experiements}
In this section, we evaluate the performance of our graph clustering and multilevel graph partitioning algorithms.
We compare ourselves against kMetis, which is probably the most widely used partitioning algorithm and the results of KaHIP presented in~\cite{pcomplexnetworksviacluster}. \\

\vspace*{-.5cm}
\paragraph{Methodology.} We implemented our algorithms using C++. Our implementation uses the 
\begin{wraptable}{r}{.35\textwidth}
\scriptsize
\centering

\vspace*{-.5cm}
\caption{Benchmark set}
\label{tab:scalefreegraphstable}
\begin{tabular}{l|r|r|r}
Graph & $n$ & $m$ & Ref.   \\
\hline
nlpkkt200    & $\approx$16.2M & $\approx$215M   & \cite{sparsematrixcollection}  \\
uk-2002      & $\approx$18.5M & $\approx$262M   & \cite{webgraphWS} \\
arabic-2005  & $\approx$22.7M & $\approx$553M   & \cite{webgraphWS}\\
nlpkkt240    & $\approx$27.9M & $\approx$373M   & \cite{sparsematrixcollection} \\
it-2004      & $\approx$41.2M & $\approx$1G     & \cite{sparsematrixcollection}\\
twitter      & $\approx$41.6M & $\approx$1.2G   & \cite{webgraphWS}\\
sk-2005      & $\approx$50.6M & $\approx$1.8G   & \cite{webgraphWS}\\
webbase-2001 & $\approx$118M  & $\approx$854M   & \cite{sparsematrixcollection}  \\
uk-2007      & $\approx$106M  & $\approx$3.3G   & \cite{webgraphWS}             \\
rgg31        & $\approx$2.1G  & $\approx$21.9G  & \cite{kappa}                 \\
\end{tabular}
\vspace*{-.5cm}
\end{wraptable}
STXXL library~\cite{STXXLWebPage} to a large extend, \ie external arrays, sorting algorithms and priority queues. All binaries were built using g++ 4.8.2. 
We  run our algorithm once in order to save running time and report cut size, running time, internal memory consumption and I/O volume. 
Experiments were run on two machines. Machine A is used in Section~\ref{s:expclustering}. It has two Intel Xeon X5550 running at 2.66 GHz (4-Core) with 48 GB RAM and 8xSATA 1 TB (read 120 MB/s, write 120 MB/s). Machine B is used in Section~\ref{s:exppartitioning}. It has two Intel Xeon E5-2650v2 running at 2.6 GHz (8-Core) with 128Gb RAM and 4xSSD 1 TB (read 1440 MB/s, write 1440 MB/s). 
The size of a block during the experiments is set to 1 MB. Using a larger block size does not yield an advantage because of the parallel prefetching read/write algorithms that are implemented within the STXXL. 

\paragraph*{Instances.}
We evaluate our algorithms on graphs collected from \cite{sparsematrixcollection,webgraphWS,kappa}. 
Table~\ref{tab:scalefreegraphstable} summarizes the main properties of the benchmark set. 
Our benchmark set includes a number of social networks and web graphs.  
\Id{rgg$X$} is a \emph{random geometric graph} from \cite{kappa} with
$2^{X}$ nodes where nodes represent random points in the unit square and edges
connect nodes whose Euclidean distance is below $0.55 \sqrt{ \ln n / n }$.
This threshold was chosen in order to ensure that the graph is almost certainly connected. 

\subsection{Graph Clustering Algorithms}
\label{s:expclustering}
We now compare different graph clustering algorithms not using a size-constraint.
We do this on the nine largest graphs from our collection (excluding \Id{rgg31}). 
By default, we perform three label propagation iterations.
We use the following algorithm abbreviations: LP - label propagation, SE - semi-external, E - external and BT for the coloring-based graph clustering algorithm. 
We use the variant of the coloring-based clustering algorithm where each bucket fits into the internal memory (this turned out to be true for all instances).
The external label propagation algorithm was run with 1 GB of internal memory. 
Figure~\ref{fig:sub1} summarizes the results (a figure illustrating the IO volume can be found in Appendix~\ref{fig:sub3}).
First of all, the semi-external and the external label propagation outperform the coloring-based clustering algorithm. 
This can be explained by the fact that each tuple in the bucket uses 20 bytes (4 bytes per element of the tuple). 
Hence, the sorting and scanning operations consume a significantly larger amount of time. 
However, the coloring-based clustering algorithm would be able to compute a graph clustering fulfilling a size-constraint in the external memory model. 
We also run LP with the active nodes strategy (see Appendix for details on the active nodes algorithm). It turns out that the LP with active nodes is faster in both models as soon as enough iterations are performed. 
\begin{figure}[h]
\vspace*{-.75cm}
\begin{center}
  \hspace*{-5mm}\includegraphics[scale=0.25]{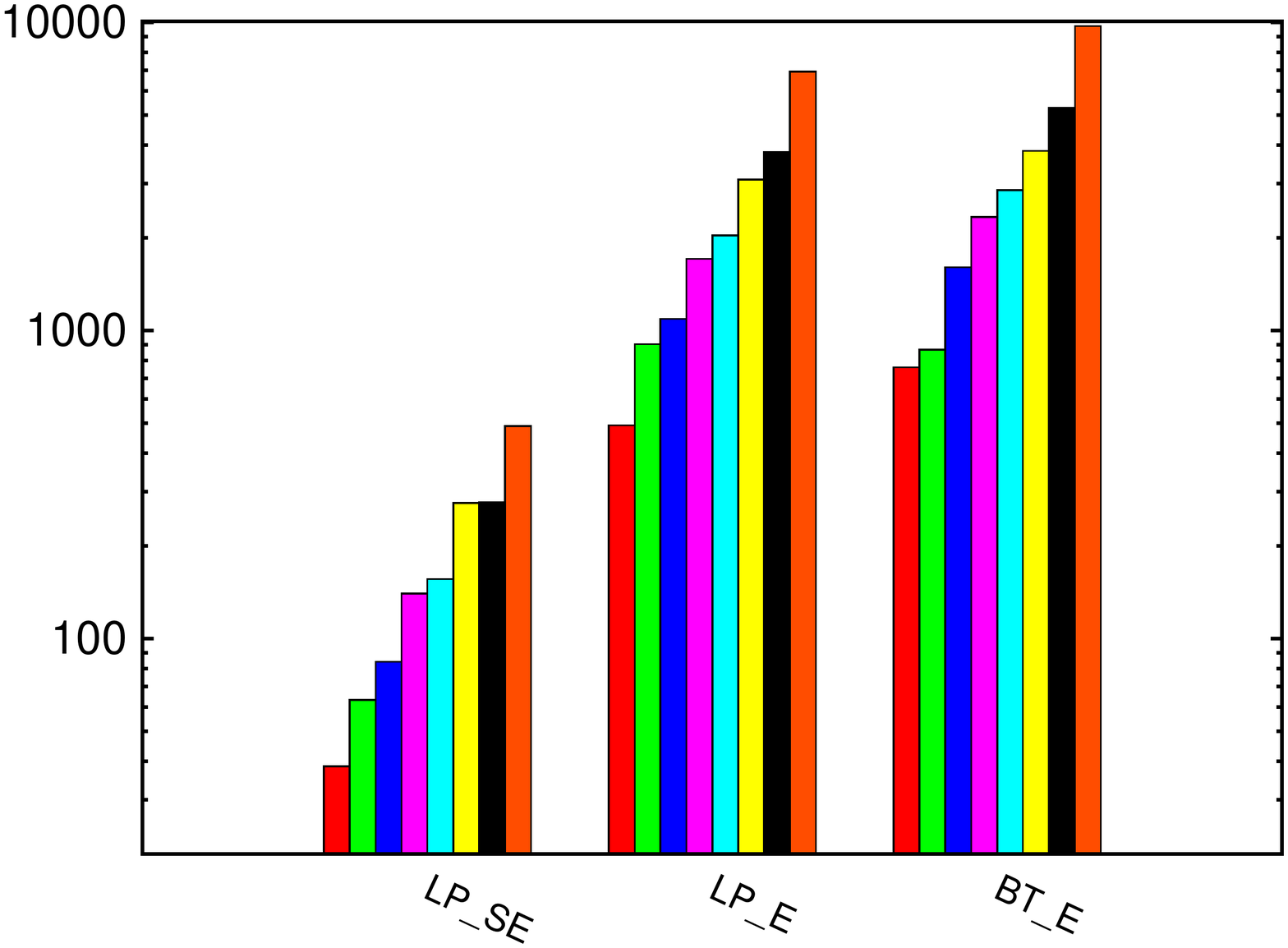}
  \includegraphics[scale=0.25]{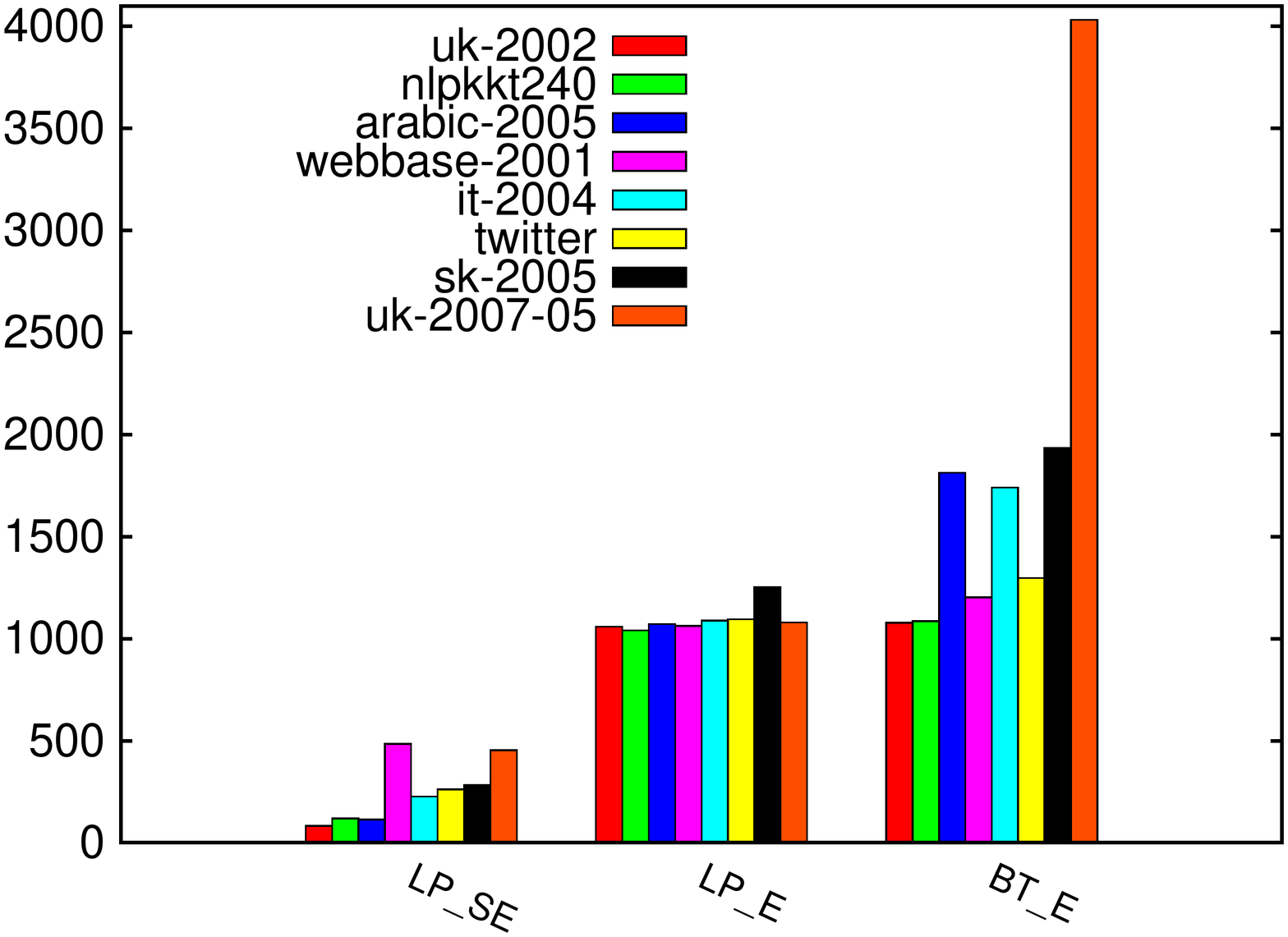}
  \vspace*{-.5cm}
  \caption{Left: Running time in seconds. Right: Memory consumption in Mb.}
  \label{fig:sub1}
  \end{center}
  \vspace*{-.5cm}
  \label{fig:sub2}
\label{fig:test}
\end{figure}

\vspace*{-1cm}
\subsection{Multilevel Graph Partitioning}
\label{s:exppartitioning}
We now present the results of the semi-external multilevel graph partitioning algorithm. 
We mainly use the semi-external label propagation algorithm with size-constraints and its parallel version to compute graph clusterings during coarsening and to perform refinement during uncoarsening. 
Our experiments focus on the four web graphs from our benchmark set that are used in \cite{pcomplexnetworksviacluster} and \Id{rgg31}. 
During coarsening we use $L_\text{max}$ as a size-constraint on the clusters. 
Note that this is an much weaker restriction on the cluster sizes than the one used in \cite{pcomplexnetworksviacluster}.
Using this weaker constraint speeds up the algorithm significantly.
The random geometric graph has been partitioned into two blocks 
\begin{wraptable}{r}{.3\textwidth}
\scriptsize
\centering
\vspace*{.5cm}
\caption{Results}
\begin{tabular}{l|rrr|}
algorithm & time[s] & cut   & mem  \\
\hline
& \multicolumn{3}{c|}{arabic-2005} \\
\hline
LP\_SE    & 81.3   & 2.04M  & 0.48GB    \\
P\_LP\_SE & 36.0 & 2.28M & 1.85GB\\
KaHIP     & 111.2   & 1.87M &   \\
kMetis    & 99.6    & 3.50M &         \\
          \hline
& \multicolumn{3}{c|}{uk-2002} \\
          \hline
LP\_SE & 57.7   & 1.46M & 0.56GB\\    
P\_LP\_SE & 39.7 & 1.54M & 1.53GB \\
KaHIP & 71.7    & 1.43M &  \\  
kMetis& 63.7    & 2.41M &         \\
          \hline
       & \multicolumn{3}{c}{sk-2005}\\
          \hline
LP\_SE & 257.7  & 19.00M & 1.59 Gb\\   
P\_LP\_SE & 203.4 & 22.25M &  7.76 Gb \\
KaHIP & 387.1   & 20.34M &\\
kMetis& 405.3   & 18.56M & \\  
          \hline
& \multicolumn{3}{c|}{uk-2007} \\
\hline
LP\_SE    & 435.1                       & 4.18M                             & 4.26GB                    \\
P\_LP\_SE & 209.1 & 4.55M &  8.58GB \\
KaHIP     & 626.6                        & 4.10M                             &  \\
kMetis    & 827.6                        & 10.86M                            &                        \\
          \hline
& \multicolumn{3}{c|}{rgg31} \\
          \hline
LP\_SE& 3803.3             & 341K & 34.40GB           \\
P\_LP\_SE & 3312.3 & 340K & 96.40GB\\
\end{tabular}
\label{tab:hugesocialresultscuts}
\vspace*{.5cm}
\end{wraptable}
\begin{wrapfigure}{r}{.4\textwidth}
\begin{center}
\vspace*{-1.25cm}
\includegraphics[width=6cm]{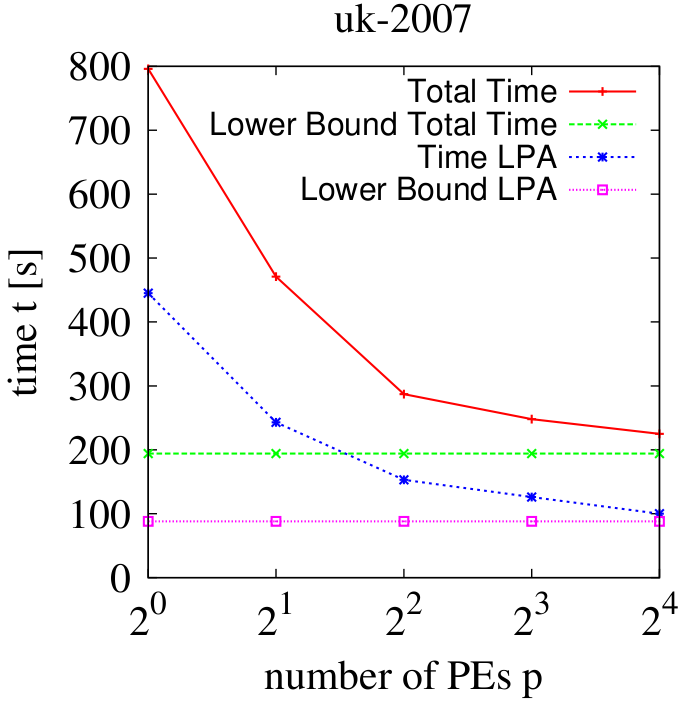}
\end{center}
\vspace*{-.5cm}
\caption{Scalability of the parallel semi-external graph partitioning algorithm on uk-2007.}
\vspace*{-.5cm}
\label{fig:parallelscalability}
\end{wrapfigure}
and the other graphs into six-teen blocks using $\epsilon=0.03$ so that we can compare ourselves with the results presented by Meyerhenke \etal \cite{pcomplexnetworksviacluster}.  The values for KaHIP and kMetis have been taken from
  their paper and the experiments have been performed on different machine with 1TB main memory. 
Table~\ref{tab:hugesocialresultscuts} summarizes the results. The parallel semi-external LPA uses 16 threads (hyperthreading) and the mem column displays how much memory has been used by our algorithm.  Our algorithm computes partitions that are almost as good as those computed by KaHIP. 
In the worst case, our algorithm cuts about 9\% more edges than KaHIP. On average we cut about 1\% more edges. 
On the other hand, our algorithm computes much better cuts than kMetis (except on the graph sk-2005). 
On average our algorithm cuts 73\% less edges then kMetis. Note that the running time of our sequential semi-external algorithm is always smaller than the running time of KaHIP. This  is partially due to the fact that we use weaker size-constraint during coarsening which makes the contracted even smaller compared to the contracted graphs computed in KaHIP and the fact that we use less label propagation rounds. 
After the first contraction step we switched to the internal memory implementation of KaHIP to partition the coarser graph. 
Figure~\ref{fig:parallelscalability} presents the scalability of our shared memory parallel semi-external algorithm on uk-2007. 
We can see that the algorithm scales well but comes with parallelization overheads. 
However, using 16 threads the parallelization always speeds up the computations compared to the sequential algorithm. 
On the other hand the parallelization gets close to the lower bound which is given by the IO that needs to be performed.
\vspace*{-.25cm}
\section{Conclusion and Future Work}
\vspace*{-.15cm}
\label{s:conclusion}
We presented algorithms that are able to partition and cluster huge complex networks with billions of edges on cheap commodity machines.
This has been achieved by using a (semi-)external variant of the size-constrained LPA that can be used for coarsening and as a simple local search algorithm.
A shared memory parallelization of the algorithm further reduced the running time of the algorithm.
Moreover, we presented the \emph{first} fully external graph clustering/partitioning algorithm that is able to deal with a size-constraint.
Our experiments indicate that our semi-external algorithms are able to compute high quality partitions in time is compareble to an efficient internal memory implementation.
As a part of future work, it might be interesting to define a (semi-)external clustering algorithm that optimizes modularity.
This can be done by using the techniques of this paper but using different update rules to compute the cluster ID of a node.
\vfill
\pagebreak
\bibliographystyle{splncs}
\bibliography{mybib,phdthesiscs}{}

\begin{thebibliography}{10}

\bibitem{kyrola2012graphchi}
Kyrola, A., Blelloch, G., Guestrin, C.:
\newblock {Graph{C}hi: Large-scale Graph Computation on Just a PC}.
\newblock In: Proc. of the 10th USENIX Symp. on Operating Systems Design and
  Implementation (OSDI). Volume~8. (2012)  31--46

\bibitem{GareyJS74some}
Garey, M.R., Johnson, D.S., Stockmeyer, L.:
\newblock Some {S}implified {NP}-complete {P}roblems.
\newblock (1974)  47--63

\bibitem{BuiJ92}
Bui, T.N., Jones, C.:
\newblock {Finding Good Approximate Vertex and Edge Partitions is {N}{P}-Hard}.
\newblock Information Processing Letters \textbf{42}(3) (1992)  153--159

\bibitem{journals/algorithmica/AbelloBW02}
Abello, J., Buchsbaum, A.L., Westbrook, J.:
\newblock {A Functional Approach to External Graph Algorithms}.
\newblock Algorithmica \textbf{32}(3) (2002)  437--458

\bibitem{GPOverviewBook}
Bichot, C., Siarry, P., eds.:
\newblock Graph Partitioning.
\newblock Wiley (2011)

\bibitem{SPPGPOverviewPaper}
Bulu\c{c}, A., Meyerhenke, H., Safro, I., Sanders, P., Schulz, C.:
\newblock {Recent Advances in Graph Partitioning}.
\newblock Technical Report ArXiv:1311.3144 (2013)

\bibitem{diekmann2000shape}
Diekmann, R., Preis, R., Schlimbach, F., Walshaw, C.:
\newblock {Shape-optimized Mesh Partitioning and Load Balancing for Parallel
  Adaptive FEM}.
\newblock Parallel Computing \textbf{26}(12) (2000)  1555--1581

\bibitem{Karypis06}
Abou-Rjeili, A., Karypis, G.:
\newblock {Multilevel Algorithms for Partitioning Power-Law Graphs}.
\newblock In: Proc. of 20th Int. Parallel and Distributed Processing Symp.
  (2006)

\bibitem{meyerhenke2006accelerating}
Meyerhenke, H., Monien, B., Schamberger, S.:
\newblock {Accelerating Shape Optimizing Load Balancing for Parallel FEM
  Simulations by Algebraic Multigrid}.
\newblock In: Proc. of 20th Int. Parallel and Distributed Processing Symp.
  (2006)

\bibitem{chaco}
Hendrickson, B.:
\newblock {Chaco: Software for Partitioning Graphs}.
\newblock {\protect\url{http://www.cs.sandia.gov/~bahendr/chaco.html}}

\bibitem{walshaw2000mpm}
Walshaw, C., Cross, M.:
\newblock {Mesh Partitioning: A Multilevel Balancing and Refinement Algorithm}.
\newblock SIAM J. on Scientific Computing \textbf{22}(1) (2000)  63--80

\bibitem{karypis1998fast}
Karypis, G., Kumar, V.:
\newblock {A Fast and High Quality Multilevel Scheme for Partitioning Irregular
  Graphs}.
\newblock SIAM J. on Scientific Computing \textbf{20}(1) (1998)  359--392

\bibitem{helpfulsetsinpractice}
Monien, B., Schamberger, S.:
\newblock {Graph Partitioning with the Party Library: Helpful-Sets in Practice}
  (2004)

\bibitem{Scotch}
Pellegrini, F.:
\newblock {Scotch Home Page}.
\newblock {\protect\url{http://www. labri.fr/pelegrin/scotch}}

\bibitem{labelpropagationclustering}
Raghavan, U.N., Albert, R., Kumara, S.:
\newblock {Near Linear Time Algorithm to Detect Community Structures in
  Large-Scale Networks}.
\newblock Physical Review E \textbf{76}(3) (2007)

\bibitem{pcomplexnetworksviacluster}
Meyerhenke, H., Sanders, P., Schulz, C.:
\newblock {Partitioning Complex Networks via Size-constrained Clustering}.
\newblock In: Proc. of the 13th Int. Symp. on Experimental Algorithms (SEA'14).
  LNCS, Springer (2014)

\bibitem{Stanton:2012:SGP:2339530.2339722}
Stanton, I., Kliot, G.:
\newblock {Streaming Graph Partitioning for Large Distributed Graphs}.
\newblock In: Proc. of the 18th ACM SIGKDD Int. Conf. on Knowledge Discovery
  and Data Mining. KDD '12, ACM (2012)  1222--1230

\bibitem{kabapeE}
Sanders, P., Schulz, C.:
\newblock {Think Locally, Act Globally: Highly Balanced Graph Partitioning}.
\newblock In: Proc. of the 12th Int. Symp. on Experimental Algorithms (SEA'13).
  LNCS, Springer (2013)

\bibitem{ZehIO}
Zeh, N.:
\newblock {I/O-efficient Graph Algorithms}.
\newblock EEF Summer School on Massive Data Sets (2002)

\bibitem{Sanders99fastpriority}
Sanders, P.:
\newblock Fast priority queues for cached memory.
\newblock ACM Journal of Experimental Algorithmics \textbf{5} (1999)  312--327

\bibitem{journals/algorithmica/Arge03}
Arge, L.:
\newblock {The Buffer Tree: A Technique for Designing Batched External Data
  Structures.}
\newblock Algorithmica \textbf{37}(1) (2003)  1--24

\bibitem{sparsematrixcollection}
{University of Florida. Sparse Matrix Collection.}

\bibitem{webgraphWS}
{Laboratory of Web Algorithms, University of Milano, Datasets}

\bibitem{kappa}
Holtgrewe, M., Sanders, P., Schulz, C.:
\newblock {Engineering a Scalable High Quality Graph Partitioner}.
\newblock Proc. of the 24th Int. Parallel and Distributed Processing Symp.
  (2010)  1--12

\bibitem{STXXLWebPage}
Beckmann, A., Bingmann, T., Dementiev, R., Sanders, P., Singler, J.:
\newblock {STXXL Home Page}.
\newblock {\url{http://stxxl.sourceforge.net/}}

\end{thebibliography}

\newpage
\begin{appendix}
\section{Omitted Proofs of the Theorems and Lemmas}

\subsection{Lemma \ref{th:bucket_aux}}
\begin{proof}
\label{th:bucket_aux_proof}
For sorting the bucket, we need $\mbox{Sort}(|\mathcal{T}|)$ I/O operations. We need $\mbox{Scan}(|\mathcal{T}|)$ I/O operations for scanning the bucket. Hence, the algorithm uses $\mbox{Sort}(|\mathcal{T}|)$ I/O operations. \qed
\end{proof}

\subsection{Theorem \ref{th:bucket_main}}
\begin{proof}
\label{th:bucket_main_proof}
Adding the information about the colors of target nodes to the edges requires $\mbox{Sort}(|E|)$ I/O operations. Our bucket initialization uses $\mbox{Scan}(|E|)$ I/O operations. Suppose we have the buckets $\mathcal{T}_1, \ldots, \mathcal{T}_\ell$. All buckets can be processed using $\mbox{Sort}(|\mathcal{T}_1|) + \ldots + \mbox{Sort}(|\mathcal{T}_\ell|) = \mbox{Sort}(|E|)$ I/O operations. Overall, we use $\mbox{Scan}(|E|)$ I/O operations for pushing tuples with the new cluster IDs into the respective buckets. Hence, the algorithm can be implemented using $\mbox{Sort}(|E|)$ I/O operations. \qed

\end{proof}

\subsection{Lemma \ref{th:bucket_blubb}}
\begin{proof}
Consider a cluster ID $c$ and let the list $N_c$ be $\{v_1, v_2, \dots, v_i, v_{i+1}, \dots, v_k\}$. 
In the beginning, there is triple $(v_1, c, size)$ in the priority queue with the actual size of the cluster $c$ due way the priority queue is initialized. 
When we process $v_i$, we put the triple with the updated size into the priority queue for \emph{the} pair $(v_{i+1}, c)$ in the list $M_{v_i}$. Hence, when we process $v_{i+1}$ the size of the cluster will also be up-to-date (since it is the next adjacent node of cluster $c$ being processed).
When $v$ is processed the triples with key value $v$ are on top of the priority queue, due to the increasing processing order.
\end{proof}

\subsection{Theorem \ref{th:bucket_sz}}
\begin{proof}
\label{th:bucket_sz_proof}
First, we prove the complexity to create the data structure containing the clusters sizes. 
In the worst case, each tuple $(v, \mbox{cluster}[u], u, w(v, u), \mbox{color}[u])$ of the bucket $\mathcal{T}$ has a unique cluster ID and also the cluster IDs of each node $v$ is unique. 
Hence, we need an additional amount of $\mathcal{O}(|\mathcal{T}|)$ internal memory for the hash table. 
For the saving and sorting the cluster IDs from the bucket we use $\mathcal{O}(1)$ I/O operations since we have the bucket in internal memory. Reading and writing the sizes of the clusters to and from the external array uses $\mbox{Scan}(|V|)$ I/O-operations.

Now we estimate the amount of buckets that fit into internal memory. 
There are two cases. 
If a bucket does not fit into internal memory, we need to divide it into multiple buckets. 
Since the overall size of all buckets is $\mathcal{O}(|E|)$ the minimum amount of buckets (such that each fits into internal memory) is $\mathcal{O}(\frac{|E|}{M})$. 
Otherwise, if all buckets fit into internal memory, we have $|C|$ buckets. 
Since we want each bucket to fit into internal memory, we have $\max(\frac{|E|}{M}, |C|)$ buckets.

The overall I/O-volume is estimated as follows: for each bucket we need $\mbox{Scan}(|V|)$ additional I/O-operations to create the data structure containing the sizes of the clusters. There are at most $\max(\frac{|E|}{M}, |C|)$ buckets. Hence, the total number of I/O-operations is $\max(\frac{|E|}{M}, |C|) \cdot \mbox{Scan}(|V|) + \mbox{Sort}(|E|)$ (to perform the main part of the bucket clustering algorithm).
\end{proof}

\subsection{Lemma \ref{th:bucket_lemma_sz_ext}}
\begin{proof}
\label{th:bucket_lemma_sz_ext_proof}

We need $\mbox{Sort}(|\mathcal{T}|)$ I/O-operations to sort the bucket and to build the lists $N$ and $\mathcal{M}$. The operations $\proc{pop}$ and $\proc{push}$ of the priority queue have amortized $\frac{1}{B}\log_{\frac{M}{B}}{\frac{N}{B}}$ cost. The number of $\proc{push}$ (or $\proc{pop}$) operations is equal to $|\mathcal{M}|$. This is due to the fact that each element of this list means that two nodes in the bucket have to take the size of the same cluster into account.
Thus, we need to forward the information from the first to the second node. This means that the total cost of all operations is $\mbox{Sort}(|\mathcal{T}|)$. 
Iterating through the bucket costs $\mbox{Scan}(|\mathcal{T}|)$. Hence, processing a bucket costs $\mbox{Sort}(|\mathcal{T}|)$ in total.
\end{proof}

\subsection{Theorem \ref{th:bucket_th_sz_ext}}
\begin{proof}
\label{th:bucket_th_sz_ext_proof}
We have $|C|$ buckets. Processing a bucket costs $\mbox{Sort}(|\mathcal{T}|)$ I/Os. The overall number of elements in the buckets is $\mathcal{O}(|E|)$. Hence, processing all buckets takes $\mbox{Sort}(|E|)$ I/O-operations. After we processed a bucket, we update the sizes of the clusters that changed during processing. This takes $\mbox{Scan}(|V|)$ I/O-operations. Thus, one iteration of the bucket clustering algorithm costs $\mbox{Sort}(|E|) + |C| \cdot \mbox{Scan}(|V|)$ I/O-operations.
\end{proof}

\section{Active Node Approach}
\label{v:active_node_approach}
The active nodes approach can be used to speed up computations of LP~\cite{pcomplexnetworksviacluster}. 
The main idea is that after the first round of the algorithm, a node can only change its cluster if one or more of its neighbors changed its cluster in the previous round. 
The active node approach keeps track of nodes that can potentially change their cluster. 
A node is called \emph{active} if at least one of its neighbors changed its cluster in the \emph{previous round}.

To translate this into our computational models, we additionally keep the set of active nodes and calculate a new cluster ID only for these nodes. 
More precisely, in the beginning all nodes are active. 
After each round the algorithm updates the set of active nodes, \ie it inserts the neighbors of nodes which have changed their cluster and deletes nodes whose neighbors have not changed their cluster. 
In the worst case the computational complexity of the LPA remains the same. 
However, our experiments show that the active node heuristic decreases the computational time of the LP algorithm.

To implement the label propagation algorithm with active nodes, we use two priority queues (external or internal, depending of the computational model). 
Both priority queues contain active nodes: the first priority queue contains active nodes that will be processed in the current round and the second priority queue contains nodes which will be processed in the next round. 
If some node $v$ changes its cluster then we push all its neighbors having a smaller ID to first priority queue and and the neighbors having a larger ID to the second priority queue. 
We use the priority queues because the algorithm must process the nodes in increasing order of their IDs. 
In the external memory model,  we store for each edge $(v, u)$  in the array of edges additional information about the cluster of $u$. 
This allows us to detect if the cluster of node has changed. 
We add $\mbox{cluster}[u]$ in the array of edges in the beginning of algorithm using $\mbox{Sort}(|E|)$ I/O operations and maintain them up-to-date during the course of the algorithm.

\paragraph*{Experiments.}
All experiments using the active nodes strategy have been performed on a machine having 2 x Intel Xeon X5355 2.66 GHz (4-Core), 24Gb RAM, 4x SATA 250GB.
Table~\ref{tab:active_node_approach} shows that the active node approach can speed up computations if the number of label propagation iterations is large.
\begin{table}[h]
\scriptsize
\centering
\vspace*{-.5cm}
\caption{Comparison of LP and LP with active nodes strategy on graph \Id{rgg22}.}
\begin{tabular}{l|r r r r}
\multirow{2}{*}{algorithm}          & \multicolumn{4}{c}{Rounds} \\
\cline{2-5}
& 3 & 4 & 5 & 7\\
\hline
LP\_SE & \textbf{34.5} s & 47.2 s & \textbf{54.7 s} & 76.8 s\\
LP\_SE\_A & 37.1 s & \textbf{44.6} s & 55.9 s  & \textbf{69.2} s \\
\hline

\multirow{2}{*}{algorithm} & \multicolumn{4}{c}{Rounds}\\
\cline{2-5}
& 11 & 12 & 13 & 15\\
LP\_E & \textbf{350.6} s & \textbf{380.4} s & 410.9 s & 471.4 s \\
LP\_E\_A & 380.8 s & 383.6 s & \textbf{384.1} s & \textbf{388.1} s\\

\end{tabular}
\label{tab:active_node_approach}
\end{table}

\section{Additional Figures}
\begin{figure}
\centering
\vspace*{-8mm}
  \includegraphics[scale=0.3]{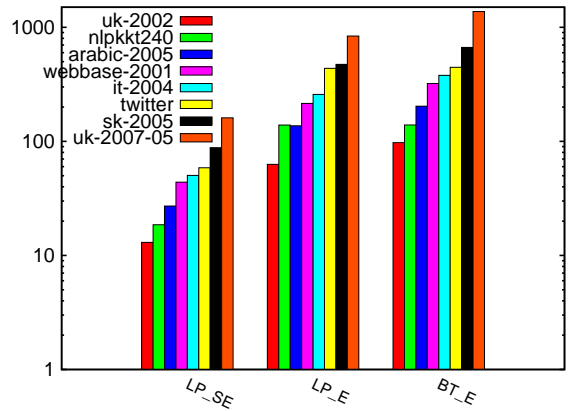}
\vspace*{-6mm}
  \caption{I/O volume of different graph clustering algorithm in GB.}
  \label{fig:sub3}
\end{figure}
\vfill
\end{appendix}

\end{document}